\documentclass[conference]{IEEEtran}
\IEEEoverridecommandlockouts
\usepackage{cite}
\usepackage{amsmath,amssymb,amsfonts}
\usepackage{amssymb}
\usepackage{algorithm}
\usepackage{algorithmicx}
\usepackage{algpseudocode}
\usepackage{amsmath}
\usepackage{graphicx}
\usepackage{textcomp}
\usepackage{xcolor}
\usepackage{epstopdf}
\usepackage{subfigure}
\usepackage{multirow}   
\usepackage{booktabs}    
\usepackage{float}
\usepackage{caption}
\def\BibTeX{{\rm B\kern-.05em{\sc i\kern-.025em b}\kern-.08em
		T\kern-.1667em\lower.7ex\hbox{E}\kern-.125emX}}

\begin{document}

\title{Beam Domain Channel Estimation for Spatial Non-Stationary Massive MIMO Systems\\	
{\footnotesize }
}

\author{Lin Hou\textsuperscript{1,2}, Hengtai Chang\textsuperscript{2,1*}, Cheng-Xiang Wang\textsuperscript{1,2*}, Jie Huang\textsuperscript{1,2}, and Songjiang Yang\textsuperscript{2}\\
	\textsuperscript{1}{National Mobile Communications Research Laboratory,}\\
	{School of Information Science and Engineering, Southeast University, Nanjing 210096, China.}\\
	\textsuperscript{2}{Purple Mountain Laboratories, Nanjing 211111, China.}\\
	\textsuperscript{*}{Corresponding Authors}\\
	\centerline{Email: hou\_lin@seu.edu.cn, changhengtai@pmlabs.com.cn,}\\
		\centerline{ \{chxwang, j\_huang\}@seu.edu.cn, yangsongjiang@pmlabs.com.cn}
}

\maketitle

\begin{abstract}
In massive multiple-input multiple-output (MIMO) systems, the channel estimation scheme is subject to the spatial non-stationarity and inevitably power leakage in the beam domain. 
In this paper, a beam domain channel estimation scheme is investigated for spatial non-stationary (SNS) massive MIMO systems considering power leakage.
Specifically, a realistic massive MIMO beam domain channel model (BDCM) is introduced to capture the spatial non-stationarity considering power leakage by introducing the illustration of visibility region (VR).
Then, a beam domain structure-based sparsity adaptive matching pursuit (BDS-SAMP) scheme is proposed based on the cross-block sparse structure and power ratio threshold of beam domain channel. 
Finally, the simulation results validate the accuracy of proposed BDS-SAMP scheme with low pilot overhead and reasonable complexity by comparing with conventional schemes.

\end{abstract}


\section{Introduction}
In recent years, massive MIMO has attracted wide attention due to the improvement in efficiency of spectrum and energy \cite{J20_VTM_CXWang_6GMM, J23_COMST_CXWang_6GVISION}.
However, the computational complexity of channel modeling and the overhead of channel estimation are challenging task because of the large number of antennas \cite{J22_SCIS_XHYou_6G,J22_TVT_CXWang_PervasiveCM}.
The BDCM, transferring the channel model from array domain into the beam domain, can reduce the computational complexity and play a critical role in the design and performance evaluation \cite{J22_TWC_JBian_mMIMOBDCM}. 
In order to fully maintain the beamforming gains of massive MIMO beam domain channels, it is essential to obtain accurate knowledge of channel state information (CSI).



Massive MIMO beam domain channels display sparsity with a small number of channel elements containing most of the energy  \cite{J02_TSP_Sayeed_BDCM}.  
Numerous channel estimation algorithms based on the compressed sensing (CS) \cite{J06_TIT_Donoho_CS}  including convex optimization algorithms \cite{J17_CL_XLin_BP}, Bayesian algorithms \cite{C12_CL_CChen_SBL}, and greedy algorithms \cite{J19_WCL_YHuang_OMP}  have been employed for massive MIMO systems by leveraging the inherent sparsity of channels.
Due to the advantages in terms of implementation simplicity and computational efficiency, greedy algorithms such as orthogonal matching pursuit (OMP), sparse adaptive matching  pursuit (SAMP) \cite{J15_TSP_ZGao_DASMP}, and block-based OMP (BOMP) \cite{J18_TVT_XMa_BOMP} have been used for massive MIMO channel estimation.
Nevertheless, the above-mentioned algorithms considered massive MIMO channels to be spatial stationary (SS) with identical sparse structure.
In practice, massive MIMO channels are SNS with some scatterers visible only to a portion of the antenna array \cite{J23_TVT_YZheng_mMIMO}, resulting in varied sparse structure over array.


In \cite{J22_WCL_JChen_SBLEstimation} and \cite{J20_CL_SHou_BMPEstimation}, the authors modified the standardized 3GPP channel model to capture the spatial non-stationarity by using birth-death process, and then performed array domain channel estimation for SNS massive MIMO systems. However, 
they ignored the inevitably impact of channel power leakage on the sparsity and estimation performance since massive MIMO channels are not ideally sparse in practice.
In \cite{C16_WCSP_XGao_SDEstimation} and \cite{J19_TSP_XGao_SDEstimation}, the channel estimation schemes based on support detection (SD) \cite{J17_TCL_XGao_SDEstimation}, e.g., adaptive SD (ASD) and successive SD, were proposed by making full use of adaptive updated or successive rectangle support structure of beam domain channel considering power leakage. 
However, these schemes were based on SS massive MIMO channels assumption, which will further underestimate the power leakage of beam domain channel and cause inaccurate results of channel estimation.

To the best of our knowledge, beam domain channel estimation for SNS massive MIMO systems is still lacking in the literature. To fill this research gap, a BDS-SAMP channel estimation scheme is proposed for SNS massive MIMO systems considering sparse structure and power leakage of beam domain channel.  
The main contributions and novelties of this paper are summarized as below.

\begin{itemize}
	\item A realistic massive MIMO BDCM is introduced to capture spatial non-stationarity by introducing the VR of different clusters. Furthermore, the beam sparse structure and power leakage effect are demonstrated.
	\item A BDS-SAMP channel estimation scheme for SNS massive MIMO systems is proposed, in which the  beam support set is iteratively obtained based on the cross-block sparse structure and power ratio threshold of beam domain channel.
	\item Simulation results verify the accuracy and robustness of proposed BDS-SAMP scheme by comparing with the traditional channel estimation schemes.
\end{itemize}


The remainder of this paper is organized as follows. 
Section~II introduces the system model including SNS massive MIMO BDCM and  problem formulation.
Based on it, Section~III analyzes the power leakage effect.
In section~IV, a BDS-SAMP channel estimation scheme for SNS massive MIMO systems is proposed. 
Section~V presents the simulation results and demonstrates the predominant performance of the proposed scheme. 
Finally, conclusions are drawn in Section~VI. 

\section{System Model}

Channel estimation schemes are highly dependent on channel models.
The realistic channel model is the basis for system performance evaluation and accurate channel estimation.
In this section, a realistic SNS massive MIMO BDCM is introduced by introducing the illustration of VR. Based on it, the beam domain channel estimation problem is formulated.

\subsection{SNS Massive MIMO BDCM}
\begin{figure}[t]
	\centering\includegraphics[width=3in]{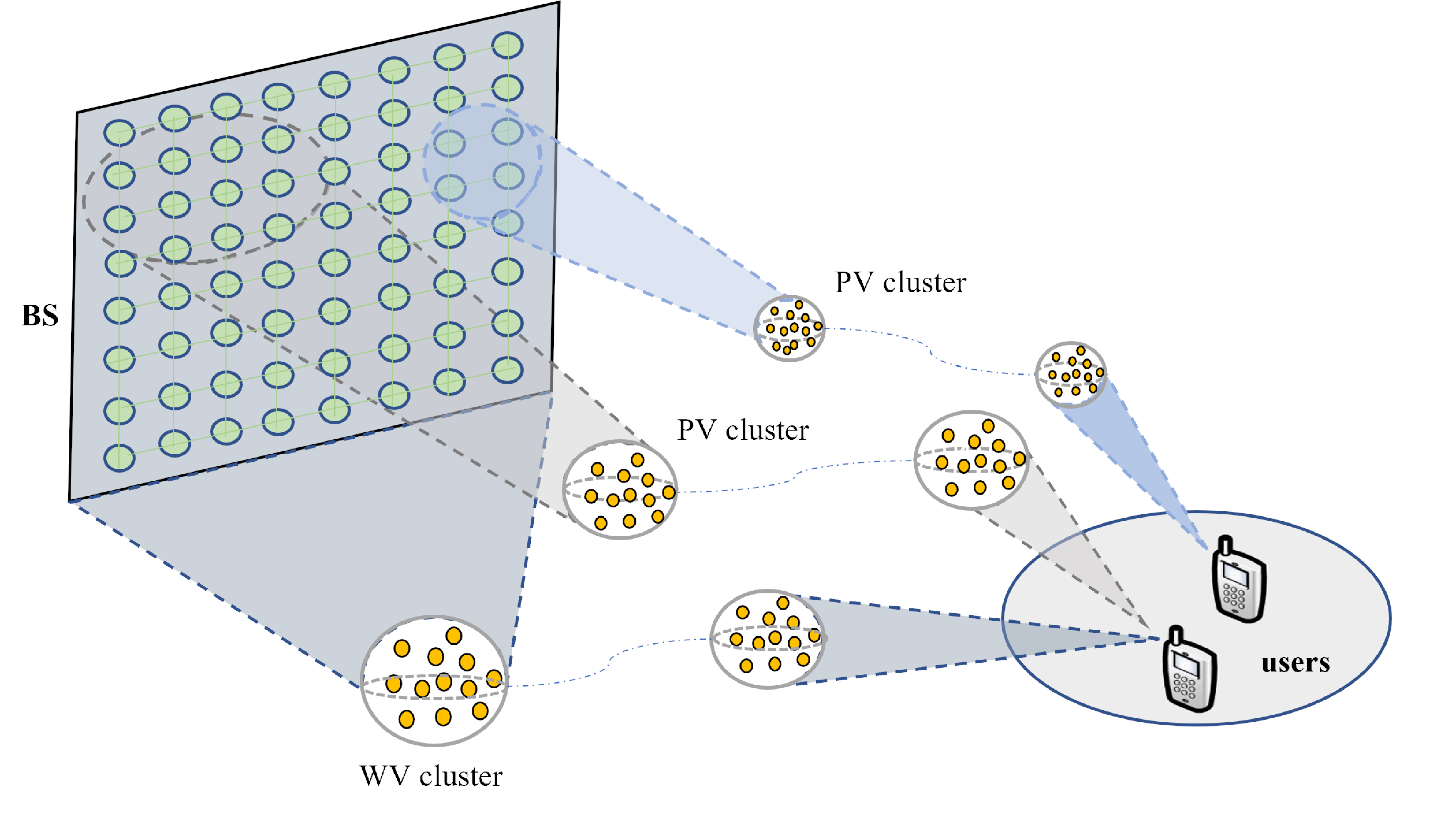}
	\caption{System structure diagram for massive MIMO systems.}
	\label{fig1}
\end{figure}

We consider a downlink massive MIMO system as shown in Fig. \ref{fig1}, where the  base station (BS) employs a large uniform planar array (UPA) with ${P}$ antennas and serves $U$ single-antenna users. 
Here, $P$ consists of ${P_v}$ antenna elements in vertical dimension multiplied by ${P_{h}}$ antenna elements in horizontal dimension.

We classify all clusters in massive MIMO systems into wholly visible (WV) clusters and partially visible (PV) clusters \cite{J22_TWC_JBian_mMIMOBDCM}.
Each cluster has a corresponding VR, the VR of WV clusters is the entire array, while that of PV clusters is the partial array. The ratio of PV clusters to total clusters is $\rho$. 
We start with the geometry-based stochastic channel model (GBSM) for massive MIMO to obtain array domain channel transfer function (CTF). The CTF matrix for the $u$th $(u=1,2,...,U)$ user can be written as
\begin{equation} 
	\begin{aligned}
		{\mathbf{H}_u} = {{\mathbf{H}}_u^{{\text{WV}}}} + {{\mathbf{H}}_u^{{\text{PV}}}} 
	\end{aligned}\label{eq1}
\end{equation}
where ${\mathbf{H}}_u^{{\text{WV}}}$ and ${\mathbf{H}}_u^{{\text{PV}}}$ are CTF matrices of WV and PV  multipath components (MPCs), which can be formulated as
\begin{equation}
	\begin{aligned} 
    {\mathbf{H}}_u^{{\text{WV}}} = 
    \sum\limits_{n \in {{\mathbf{N}}_{1}}} {\sum\limits_{m = 1}^{{M_n}} {{\beta _{n,m}}{e^{{\rm j}\psi_{n,m}}}}} {{{\mathbf{U}}\left( {\theta _{n,m}^{az},\theta _{n,m}^{el}} \right)}} 
	\end{aligned}
\end{equation}
\begin{equation}
	\begin{aligned}
    {{\mathbf{H}}_u^{{\text{PV}}}} = \sum\limits_{n \in {{\mathbf{N}}_{2}}} {\sum\limits_{m = 1}^{{M_n}} {{\beta _{n,m}}{e^{{\rm j}\psi_{n,m}}}}} {{\mathbf{\hat U}}\left( {\theta _{n,m}^{az},\theta _{n,m}^{el}} \right)}
	\end{aligned}
\end{equation}
with 
\begin{equation} 
    \psi_{n,m} =  { - 2\pi f{\tau _{n,m}} + {\varphi _{n,m}}}.
\end{equation}
Here,  $f$ indicates the carrier frequency, $M_n$ stands for the  paths number of the $n$th cluster, ${\bf{N}}_{1/2}$ are the sets of the WV/PV clusters, $\varphi _{n,m}$,  $\tau _{n,m}$ and ${\beta_{n,m}}$ denote the random initial phase, delay and coefficient of $m$th path in the $n$th cluster, respectively. 
To avoid redundancy, abbreviate all variables whose subscript is (${n,m}$) as that of $m_n$th path in the following part.
The UPA steering matrix for WV clusters is defined as ${\bf{U}} \left( {\theta _{n,m}^{az},\theta _{n,m}^{el}} \right) = {\left[ {{\mathbf{ b}}\left( {\theta _{n,m}^{el}} \right) \otimes {\mathbf{ a}}\left( {\theta _{n,m}^{az}} \right)} \right]}^{\text{T}}$, where ${\mathbf{a}}\left( {\theta _{n,m}^{az}} \right) = {\left[ {1,{e^{{\rm j}2\pi \theta _{n,m}^{az}}}, \cdots ,{e^{{\rm j}2\pi \left( {{P_h} - 1} \right)\theta _{n,m}^{az}}}} \right]^{\text{T}}}$ and  ${\mathbf{b}}\left( {\theta _{n,m}^{el}} \right) = {\left[ {1,{e^{{\rm j}2\pi \theta _{n,m}^{el}}}, \cdots ,{e^{{\rm j}2\pi \left( {{P_v} - 1} \right)\theta _{n,m}^{el}}}} \right]^{\text{T}}}$
with $\theta _{n,m}^{az}$ and $\theta _{n,m}^{el}$ standing for the spatial frequencies in azimuth and elevation directions associated with the $m_n$th path.
Without loss of generality, we usually set adjacent antennas spacing as half wavelength, thus $\theta _{n,m}^{az/el} \in \left[ { - 0.5,0.5} \right]$.
Furthermore, the steering matrix for PV clusters can be obtained via ${\bf{\hat U}}\left( {\theta _{n,m}^{az},\theta _{n,m}^{el}} \right) = {\bf{U}}\left( {\theta _{n,m}^{az},\theta _{n,m}^{el}} \right)  \odot {{\mathbf{\xi }}_{n,m}}$. The VR of $m_n$th path is defined by the matrix $\mathbf{\xi }_{n,m}$ consisting only of 0 and 1, which indicates the visibility of cluster to antenna array. 

\begin{figure}[t]
	\centering\includegraphics[width=3.5in]{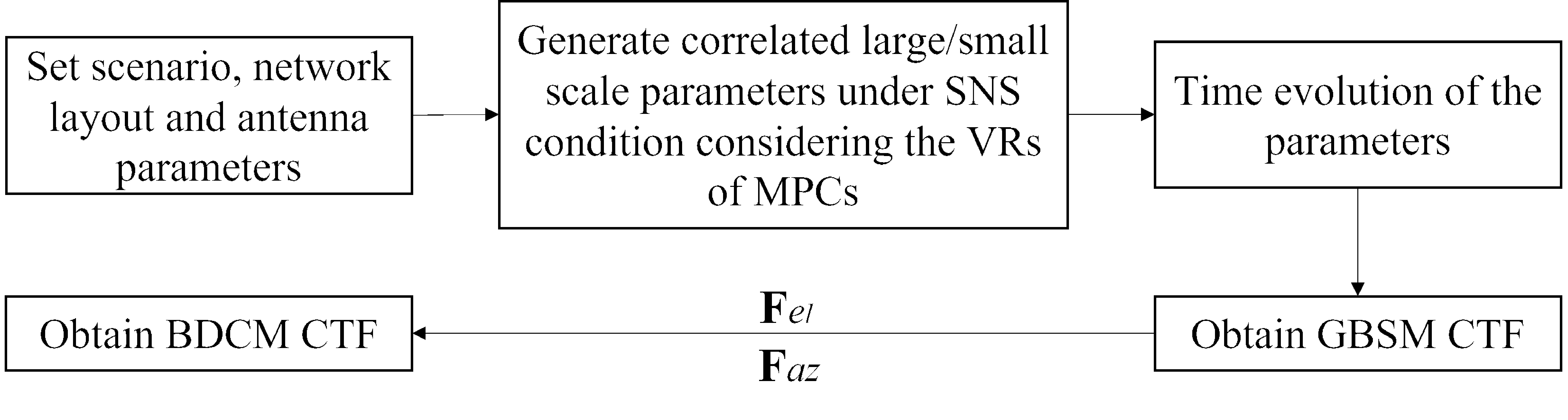}
	\caption{Flowchart of generating the SNS BDCM.}
	\label{fig2}
\end{figure}
The BDCMs can be obtained by unitary transformation of GBSMs \cite{J23_TWC_HChang_BDCM}. 
For clarity, the brief flowchart of generating the SNS BDCM is shown in Fig. \ref{fig2}. 
The array domain CTF $\textbf{H}_u$ can be converted to the beam domain CTF $\textbf{H}{_{\text B,u}}$ by two-dimensional DFT processing as follows
\begin{equation}
	\begin{aligned}
		{{\bf{H}}_{\text B,u}} = {\bf{F}}_{^{el}}^*{\bf{H}}_u {\bf{F}}_{^{az}}^{\text{T}}
	\end{aligned}\label{eq2}
\end{equation}	
with 
\begin{equation}
	\begin{aligned}
		{\bf{F}}_{^{el}} = \frac{1}{{\sqrt {{P_v}} }}\left[ {\mathbf{b}}\left( {\tilde \theta _{i}^{el}} \right) \right]_{i = 1,2,...,{P_v}}^{\text{T}}
	\end{aligned}
\end{equation}
\begin{equation}
	\begin{aligned}
		{\bf{F}}_{^{az}} = \frac{1}{{\sqrt {{P_h}} }}\left[ {\mathbf{a}}\left( {\tilde \theta _{j}^{az}} \right) \right]_{j = 1,2,...,{P_h}}^{\text{T}}
	\end{aligned}
\end{equation}
where $\tilde \theta _i^{el}$ and $\tilde \theta _j^{az}$ are the predetermined elevation and azimuth spatial frequencies of beam domain, and \textbf{F${_{el}}$} and \textbf{F${_{az}}$} represent the elevation and azimuth beamforming matrices, respectively. 
By substituting (\ref{eq1}) into (\ref{eq2}), the ${\mathbf{H}_{\text B,u}}$ can be rewritten as
\begin{equation} 
	\begin{aligned}
		{\mathbf{H}_{\text B,u}} = {{\mathbf{H}}_{\text B,u}^{{\text{WV}}}} + {{\mathbf{H}}_{\text B,u}^{{\text{PV}}}}. 
	\end{aligned}\label{eq}
\end{equation}
where ${{\mathbf{H}}_{\text B,u}^{{\text{WV}}}}$ and ${{\mathbf{H}}_{\text B,u}^{{\text{PV}}}}$ are beam domain CTF matrices of WV and PV  MPCs.

\subsection{Problem Formulation}
This paper adopts the orthogonal pilot transmission strategy and mainly focus on the downlink. Suppose BS repeatedly sends pilot sequences with the length of $U$ to different users for $Q$ times, and the channel experiences the same fading during  $K = U \times Q$ time slots.
The pilot sequence sent from BS to the users forms the pilot matrix ${\bf{X}} \in {\mathbb{C}^{U \times U}}$.
When transmitting the $q$th $(q = 1,2,..., Q)$ pilot sequence, the BS employs an analog precoder ${{\bf{F}}_q} \in {\mathbb{C}^{U \times P}}$, and the  received signals by the $u$th user can be presented by 
\begin{equation} 
	\begin{aligned}
		{{\bf{r}}_{u,q}} = {\bf{X}}{{\bf{F}}_q}{{\bf{h}}_{u}} + {{\bf{n}}_{u,q}}
	\end{aligned}
\end{equation}		
where 
${{\bf{h}}_{u}} \in {\mathbb{C}^{P \times 1}} $ is the channel vector rearranged by ${{\bf{H}}_{u}}$,
${{\bf{n}}_{u,q}} \in {\mathbb{C}^{U \times 1}} $ is the additive
white Gaussian noise vector satisfying ${{\bf{n}}_{u,q}} \sim {\cal C}{\cal N}(0,{\sigma ^2}{{\bf{I}}_U})$ with $\sigma^2$ standing for the noise power. Due to the orthogonality of ${\bf{X}}$, i.e., ${\bf{X}}{{\bf{X}}^{\rm{H}}} = {{\rm I}_U}$, we can obtain 
\begin{equation} 
	\begin{aligned}
		{{\bf{y}}_{u,q}} = {{\bf{X}}^{\rm{H}}}{{\bf{r}}_{u,q}} = {{\bf{F}}_q}{{\bf{h}}_{u}} + {{{\bf{\tilde n}}}_{u,q}}\\
	\end{aligned}\label{equ_y}
\end{equation}	
where ${{{\bf{\tilde n}}}_{u,q}} = {{\bf{X}}^{\rm{H}}}{{\bf{n}}_{u,q}}$. Define ${\bf{\tilde U}} = {{\bf{F}}_{el}} \otimes {{\bf{F}}_{az}} \in {\mathbb{C}^{P \times P }}$, ${{\bf{y}}_{u,q}}$ in (\ref{equ_y}) can be rewritten as  
\begin{equation} 
	\begin{aligned}
		{{\bf{y}}_{u,q}}  = {{\bf{\Phi }}_q}{{\bf{h}}_{\text B,u}} + {{{\bf{\tilde n}}}_{u,q}}
	\end{aligned}
\end{equation}	
where ${{\bf{\Phi }}_q} = {{\bf{F}}_q}{{{\bf{\tilde U}}}^{\rm{H}}}$, and ${{\bf{h}}_{\text B,u}} \in {\mathbb{C}^{P \times 1}} $ is the channel vector rearranged by ${{\bf{H}}_{\text B,u}}$ satisfying ${{\bf{h}}_{\text B,u}} =  {{{\bf{\tilde U}}}}{{\bf{h}}_{u}}$ according to (\ref{eq}). After $Q$ times repetitive pilot transmission, the received signal can be written as 
\begin{equation} 
	\begin{aligned}
		{{\bf{y}}_u} = {\bf{\Phi }}{{\bf{h}}_{\text B,u}} + {{{\bf{\tilde n}}}_u}
	\end{aligned}\label{equ_cs}
\end{equation}	
where ${{\bf{y}}_u} \buildrel \Delta \over = {\left[ {{\bf{y}}_{u,1}^{\rm{T}},{\bf{y}}_{u,2}^{\rm{T}},...,{\bf{y}}_{u,Q}^{\rm{T}}} \right]^{\rm{T}}} \in {\mathbb{C}^{K \times 1}}$, ${\bf{\Phi }} \buildrel \Delta \over = {\left[ {{\bf{\Phi }}_1^{\rm{T}},{\bf{\Phi }}_2^{\rm{T}},...,{\bf{\Phi }}_Q^{\rm{T}}} \right]^{\rm{T}}} \in {\mathbb{C}^{K \times P}} $ represents the measurement matrix, and ${{{\bf{\tilde n}}}_u} \buildrel \Delta \over = {\left[ {{\bf{\tilde n}}{{_{u,1}^{\rm{T}}}},{\bf{\tilde n}}{{_{u,2}^{\rm{T}}}},...,{\bf{\tilde n}}{{_{u,Q}^{\rm{T}}}}} \right]^{\rm{T}}} \in {\mathbb{C}^{K \times 1}} $.
Owing to the sparsity of vector ${\bf{h}}_{\text B,u}$, (\ref{equ_cs}) can be regarded as sparse signal recovery problem. 
However, massive MIMO channels are not ideally sparse in practice due to the power leakage, which challenges the channel recovery performance.
To better solve the problem of beam domain channel estimation, the power leakage effect is detailed introduced in next section.

\section{Analysis of Power Leakage }
\begin{figure}[t]
	\centering\includegraphics[width=3in]{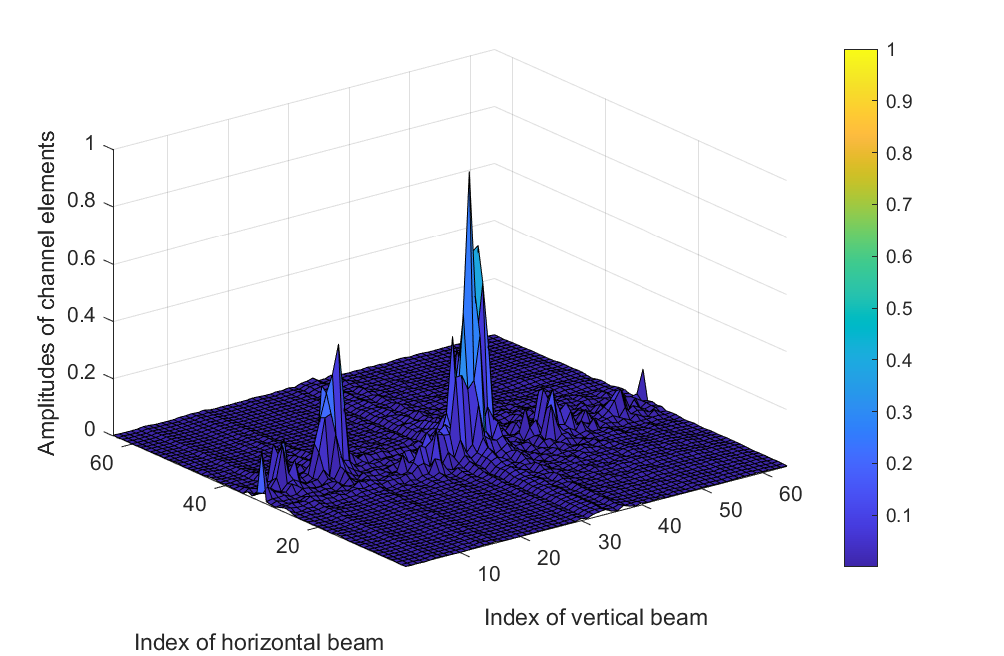}
	\caption{Sparse structure of SNS massive MIMO beam domain channel.}
	\label{fig3}
\end{figure}

The sparse structure of SNS massive MIMO beam domain channel is visualized in Fig. \ref{fig3}. It is notable that the channel amplitudes are only concentrated within a small part of the beams and the dominant beams form the shape of cross-blocks. However, it is not ideally sparse. The fact that the power of small nonzero amplitudes spreads over other beam directions refers to power leakage. In massive MIMO beam domain channels, power leakage inevitably exists between beams due to the limited beamspace resolution of small VRs and imperfect beam sampling \cite{J22_TWC_JBian_mMIMOBDCM}. 
\begin{figure}[t]
	\centering\includegraphics[width=3in]{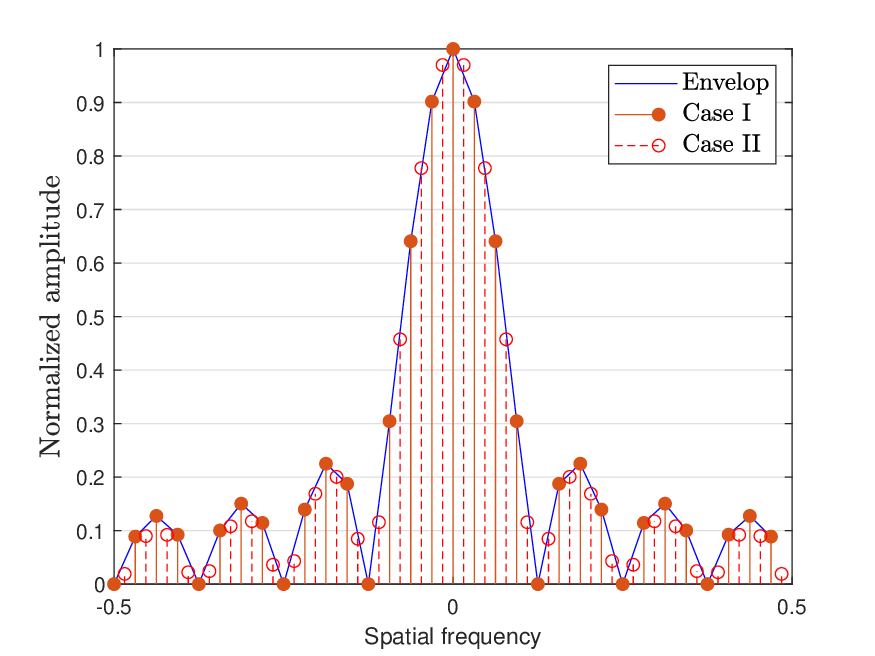}
	\caption{Envelop of ${f_{I_{s}^h,I_{e}^h}} \left( {\theta_0^{az} - \tilde \theta _j^{az}} \right)$ ($\theta_0^{az} = 0, I_{s}^h = P_h/4, I_{e}^h = P_h/2-1$).}
	\label{fig4}
\end{figure}
The beam domain channel representation of the $(i,j)$th beam  between the BS and $u$th user can be formulated as 
\begin{equation}
	\begin{aligned}
		{{\bf{h}}_{\text B,u(i,j)}} = {{\mathbf{h}}_{\text B,u(i,j)}^{{\text{WV}}}} +  {{\mathbf{h}}_{\text B,u(i,j)}^{{\text{PV}}}}
	\end{aligned}
\end{equation}	
where
\begin{equation}
	\begin{aligned}
		\begin{gathered}
			{{\mathbf{h}}_{\text B,u(i,j)}^{{\text{WV}}}} = \frac{1}{{\sqrt P }}\sum\limits_{n \in {{\mathbf{N}}_1}}^{} {\sum\limits_{m = 1}^{{M_n}} {{\beta _{n,m}}{e^{{\rm j} \psi_{n,m}}}} }  \hfill \\
			\times {f_{1,{P_v}}}\left( {\theta _{n,m}^{el} - \tilde \theta _i^{el}} \right){f_{1,{P_h}}}\left( {\theta _{n,m}^{az} - \tilde \theta _j^{az}} \right) \hfill \\ 
		\end{gathered} 
	\end{aligned}
\end{equation}	
\begin{equation}
	\begin{aligned}
		\begin{gathered}
			{{\mathbf{h}}_{\text B,u(i,j)}^{{\text{PV}}}} = \frac{1}{{\sqrt P }}\sum\limits_{n \in {{\mathbf{N}}_2}}^{} {\sum\limits_{m = 1}^{{M_n}} {{\beta _{n,m}}{e^{{\rm j} \psi_{n,m}}}} }  \hfill \\
			\times {f_{I_{s,nm}^v,I_{e,nm}^v}}\left( {\theta _{n,m}^{el} - \tilde \theta _i^{el}} \right){f_{I_{s,nm}^h,I_{e,nm}^h}}\left( {\theta _{n,m}^{az} - \tilde \theta _j^{az}} \right) \hfill \\ 
		\end{gathered} 
	\end{aligned}
\end{equation}
with 
\begin{equation}
	\begin{aligned}
		{f_{I_s,I_e}}\left( x \right) = {e^{j\pi x({I_s} + {I_e} - 2)}}\frac{{\sin \left[ {\pi x({I_e} - {I_s} + 1)} \right]}}{{\sin \left[ {\pi x} \right]}}
	\end{aligned}\label{eq3}
\end{equation}	
where ${I_{s,nm}^{h/v}}$ and ${I_{e,nm}^{h/v}}$ represent the start and end column/row indices of VR for the $m_n$th path, respectively. 

Note that ${f_{I_s,I_e}}\left( x \right)$ in (\ref{eq3}) is the Dirichlet sinc function. 
When $\rho=0$, the channel model degenerates to SS resulting in the narrowest beam, which is already taken into account in most beam domain channel estimate schemes. In this case, if the physical direction of the MPCs is perfectly sampled, i.e., $\tilde \theta _i^{el} \in {{\rm{\Lambda }}_1} \buildrel \Delta \over = \left\{ {i/{P_v} - 0.5,i = 1,2,...,{P_v}} \right\}$ and $\tilde \theta _j^{az} \in {{\rm{\Lambda }}_2} \buildrel \Delta \over = \left\{ {j/{P_h} - 0.5,j = 1,2,...,{P_h}} \right\}$, no power leakage will occur. However, in most cases $\tilde \theta _i^{el} \notin {{\rm{\Lambda }}_1}$ or $\tilde \theta _j^{az} \notin {{\rm{\Lambda }}_2}$, thus the WV MPCs result in power leakage, which is manifested by two peaks in the main beam of its envelope  \cite{J17_CL_WMa_BDSEstimation}.

When PV MPCs are considered in massive MIMO systems, the beam widens and the power leakage becomes complicated.
Fig. \ref{fig4} shows the normalized amplitude of ${f_{I_{s}^h,I_{e}^h}} \left( {\theta_0^{az} - \tilde \theta _j^{az}} \right)$ for PV MPCs with $\rho=1$ and ${P_h} = 32$. The envelop of ${f_{I_{s}^v,I_{e}^v}} \left( {\theta_0^{el} - \tilde \theta _i^{el}} \right)$ is similar as ${f_{I_{s}^h,I_{e}^h}} \left( {\theta_0^{az} - \tilde \theta _j^{az}} \right)$.
The horizontal length of VR is 8 continuous antenna elements. 
It can be readily seen that ${f_{I_{s}^h,I_{e}^h}} \left( {\theta _{0}^{az} - \tilde \theta _j^{az}} \right)$ are uniformly sampled at direction $\tilde \theta _j^{az}$  and are peaky around $\theta _{0}^{az}$.
Due to the partial visibility of MPCs, the power leakage can be observed inevitably under case I, i.e.,  $\tilde \theta _j^{az} \in {{\rm{\Lambda }}_2}$, which results in minimal power leakage, and case II, i.e., $\tilde \theta _j^{az} \notin {{\rm{\Lambda }}_2}$, which  maximizes the power leakage. 


\section{BDS-SAMP Channel Estimation Scheme }
Due to the sparsity and the inevitable power leakage of SNS MIMO beam domain channel, it is necessary to consider the sparse structure and power leakage effect when performing channel estimation.
In this section, a BDS-SAMP algorithm for SNS massive MIMO beam domain channel estimation is proposed, in which the beam support set is iteratively obtained based on the cross-block sparse structure and power ratio threshold of beam domain channel.
The pseudo-code of the proposed scheme is summarized in Algorithm \ref{A1}, which can be explained in detail as follows.

Define the residual vector of the $(k-1)$th iteration as ${{\mathbf{r}_{k - 1}}} \in {\mathbb{C}^{K \times  {1} }} $ and beam correlation between  residual vector and $p$th column of  measurement matrix ${\bf{\Phi}}_p$ as $c_p={{\bf{\Phi }}_p^{\rm{H}}{\mathbf{r}_{k - 1}}}$. The beam with higher $c_p$ means it is easier to be selected into the beam support set. Therefore, the column most relevant to residual vector can be obtained by
 \begin{equation} 
 	\begin{aligned}
 	{S_k} =\text{max}\left\{ {\left| c_p \right|_{p = 1}^P,s} \right\}
 	\end{aligned}\label{equsp}
 \end{equation}	
where ${S_k}$ is the preliminary beam support set.
In order to find out the dominant beam supports quickly, we have to determine the entries on the top, bottom, left, and right of ${S_k}$ based on the cross-block of beam domain channel, and denote them  as ${{\Upsilon } \in \Omega}$, $\Omega  \triangleq \{ 1,2, \ldots ,P\}$.
Define the power ratio threshold  as $\mu  \in \left( {0,1} \right]$. 
Then compute the power ratio of the strongest beam entry ${S_k}$ over all the dominant beam supports as
 \begin{equation} 
	\begin{aligned}
   \tilde \mu  = {{{{\mathbf{h}}_{\text B,u[{S_k}]}}} \mathord{\left/
		{\vphantom {{{{\mathbf{h}}_{\text B,u[{S_k}]}}} {\left( {{{\mathbf{h}}_{\text B(u)[{S_p}]}} + \sum\limits_{i \in \Upsilon } {{{\mathbf{H}}_{\text B(u)[i]}}} } \right)}}} \right.
		\kern-\nulldelimiterspace} {\left( {{{\mathbf{h}}_{\text B,u[{S_k}]}} + \sum\limits_{i \in \Upsilon } {{{\mathbf{h}}_{\text B,u[i]}}} } \right)}}
	\end{aligned}\label{equumu}
\end{equation}	
and refine preliminary beam support set ${S_k}$ via 
 \begin{equation} 
	\begin{aligned}
		{S_k} = {S_k} \cup \Upsilon 
	\end{aligned}\label{equus}
\end{equation}
with $\tilde \mu  \geqslant \mu $.  

Next, we get candidate list ${C_k}$ via an union ${C_k} = {\Omega _s} \cup {S_k}$ and then obtain final test via 
\begin{equation} 
	\begin{aligned}
		F = \text{max}\left\{ {\left| {{{\left( {{\bf{\Phi }}_p^{\rm{H}}{\bf{\Phi }}_p} \right)}^{ - 1}}{\bf{\Phi }}_p^{\rm{H}} {\mathbf{y}}_u} \right|_{p = 1}^{\text{card}({C_k})},s} \right\}.
	\end{aligned}\label{equf}
\end{equation}
The initial estimation of dominant entries of ${{\mathbf{h}}_{\text B,u}}$ is 
 \begin{equation} 
 	\begin{aligned}
 	{{\mathbf{\hat h}}}_{\text B,u} [F] = {\left( {{\bf{\Phi }}_F^{\rm{H}}{\bf{\Phi }}_F} \right)^{ - 1}}{\bf{\Phi }}_F^{\rm{H}}{\mathbf{y}_u}
 	\end{aligned}\label{equh}
 \end{equation}
which is obtained by essentially least squares (LS) estimation.
To get the refined beam support set, the residual vector should be updated via
 \begin{equation} 
	\begin{aligned}
		{{\mathbf{r}}_F} = {\mathbf{y}}_u -{\bf {\Phi}}_F{{\mathbf{\hat h}}}_{\text B,u} [F].
	\end{aligned} \label{equr}
\end{equation}
 
 \renewcommand{\algorithmicrequire}{\textbf{Input:}}
 \renewcommand{\algorithmicensure}{\textbf{Output:}}
 \begin{algorithm}[t]
 	\caption{Proposed BDS-SAMP Scheme}
 	\begin{algorithmic}[1] 
 		\Require {The received signal ${\bf y}_u$, measurement matrix ${\bf{\Phi}}$, power ratio threshold $\mu$ and step size $s$ = 1}.
 		\Ensure {Estimated beam domain channel for user $u$: ${{\bf{\hat h}}_{\text B,u}}$.}
 		\State Initialization: Initial residual vector ${{\mathbf{r}}_0} = {\mathbf{y}_u}$; Support set ${\Omega _s} = \emptyset$; Iteration index $k$  = 1.	
 		\While{$k \le K$} 	
 		\State Obtain ${S_k}$ via (\ref{equsp}); 
 		\State Obtain $\tilde \mu$  via (\ref{equumu});
 		\If {$\tilde \mu  \geqslant \mu $}
 		\State {Update ${S_k}$ via (\ref{equus});}
 		\EndIf	
 		\State Make candidate list via ${C_k} = {\Omega _s} \cup {S_k}$;
 		\State Obtain final test $F$ via (\ref{equf});
 		\State Obtain  LS estimation ${{\mathbf{\hat h}}}_{\text B,u} [F]$ via (\ref{equh}) ;
 		\State Compute residual ${{\mathbf{r}}_F} $ via (\ref{equr});
 		\If {$\left\| {{{\mathbf{r}}_F}} \right\|_2^2 < {\mathbf{r}}$}
 		\State {${\Omega _s} = F$;}
 		\State ${{\mathbf{r}}_k} = {{\mathbf{r}}_F}$;
 		\State \textbf{break};
 		\ElsIf {${{\mathbf{r}}_F} \geqslant {{\mathbf{r}}_{k - 1}}$}
 		\State $s = s + 1$;
 		\Else
 		\State {${\Omega _s} = F$;}
 		\State ${{\mathbf{r}}_k} = {{\mathbf{r}}_F}$;
 		\State $k = k + 1$;
 		\EndIf
 		
 		\EndWhile
 		\State \Return ${{\mathbf{\hat h}}}_{\text B,u}[{\Omega _s}]$  via (\ref{equhb}).	
 	\end{algorithmic} 
 	\label{A1}
 \end{algorithm}
Since noise can affect the received signal in different channel states, the stop iteration threshold parameters need to be set differently according to the channel states. 
The signal-to-noise ratio (SNR) is used to characterize the channel state, and the stop iteration threshold of algorithm under different SNR is set by \cite{J20_CL_SHou_BMPEstimation}
  \begin{equation} 
 	\begin{aligned}
 {\mathbf{r}} =	{{\left\| {\mathbf{y}}_u \right\|_2^2} \mathord{\left/{\vphantom {{\left\| {\mathbf{Y}} \right\|_2^2} {({{10}^{{\text {SNR} \mathord{\left/	{\vphantom {\text {SNR} {10}}} \right.	\kern-\nulldelimiterspace} {10}}}} + 1)}}} \right.\kern-\nulldelimiterspace} {({{10}^{{\text {SNR} \mathord{\left/	{\vphantom {\text {SNR}{10}}} \right.\kern-\nulldelimiterspace} {10}}}} + 1)}}.
 	\end{aligned} \label{eqr}
 \end{equation}
If the residual vector satisfies {$\left\| {{{\mathbf{r}}_F}} \right\|_2^2 < {\mathbf{r}}$}, then the beam support set and residual vector can be updated. Otherwise, compare the current residual vector ${{\mathbf{r}}_F}$ with the residual of the previous iteration ${{\mathbf{r}}_{k - 1}}$ to update the support set, residual vector, and number of iterations. Repeat the above steps to get beam support set ${\Omega _s}$ until all the pilot information has been used, and finally get the estimated channel via
\begin{equation} 
	\begin{aligned}
		{{\mathbf{\hat h}}}_{\text B,u} [{\Omega _s}] = {\left( {{\bf{\Phi }}_{\Omega _s}^{\rm{H}}{\bf{\Phi }}_{\Omega _s}} \right)^{ - 1}}{\bf{\Phi }}_{\Omega _s}^{\rm{H}}{\mathbf{y}_u}.
	\end{aligned}\label{equhb}
\end{equation}

Note that if the beam support refinement from step 4 to step 7 is deleted, the scheme will be reduced to the SAMP. The calculations of (\ref{equumu}) and  (\ref{equus}) are performed offline without additional computational complexity. The main complexity of the algorithm focuses on the matrix inverse operations in step 9 and 10, resulting in the computational complexity of ${\rm O}(2K^\frac{3}{2})$. In the proposed BDS-SAMP algorithm, the dimension of residual vector ${{\mathbf{r}}_k}$, i.e., $K$ is much smaller than that of the original channel vector ${{\mathbf{y}}_u}$, i.e., $P$ due to beam supports refinement considering the sparse channel structure and the influence of power leakage. Furthermore, the proposed BDS-SAMP algorithm can effectively reduce the pilot overhead.


\section{Simulation Results}
In this section, the performance of the proposed DBS-SAMP scheme is verified. Here, the normalized mean square error (NMSE) $\mathbb{E}\left\{ {\frac{{\left\| {{{\mathbf{h}}}_{\text B,u} - {{\mathbf{\hat h}}}_{\text B,u}} \right\|_2^2}}{{\left\| {{\mathbf{h}}}_{\text B,u} \right\|_2^2}}} \right\}$ is used to evaluate the performance between estimated channel ${{\mathbf{\hat h}}}_{\text B,u}$ and the true channel ${{\mathbf{h}}}_{\text B,u}$. 
The choice of measurement matrix is important to ensure that algorithm converges and the received signal contains sufficient information to accurately reconstruct the channel. Bernoulli random matrices have been proved to have lower mutual-column coherence  $\eta  = \mathop {\max }\limits_{i \ne j} \left| {{\mathbf{\Phi }}_i^ {\text H}{{\mathbf{\Phi }}_j}} \right|$ to achieve higher recovery accuracy \cite{C23_WCNC_JGholipour_CS}. Therefore, we define ${\bf{\Phi }}$ as Bernoulli random matrice, and select the elements of ${\bf{\Phi }}$ randomly from the set $\frac{1}{{\sqrt K }}\left\{ { - 1,1} \right\}$ with equal probability. The novel BDCM considering spatial non-stationarity is adopted in our simulation with different number of antennas.
The power ratio threshold $\mu=0.9$ and the carrier frequency $f=11$ GHz. The other channel parameters  are selected according to \cite{3GPP} in the urban microcellular (UMi) communication scenario.

\begin{figure}[t]
	\centering\includegraphics[width=3in]{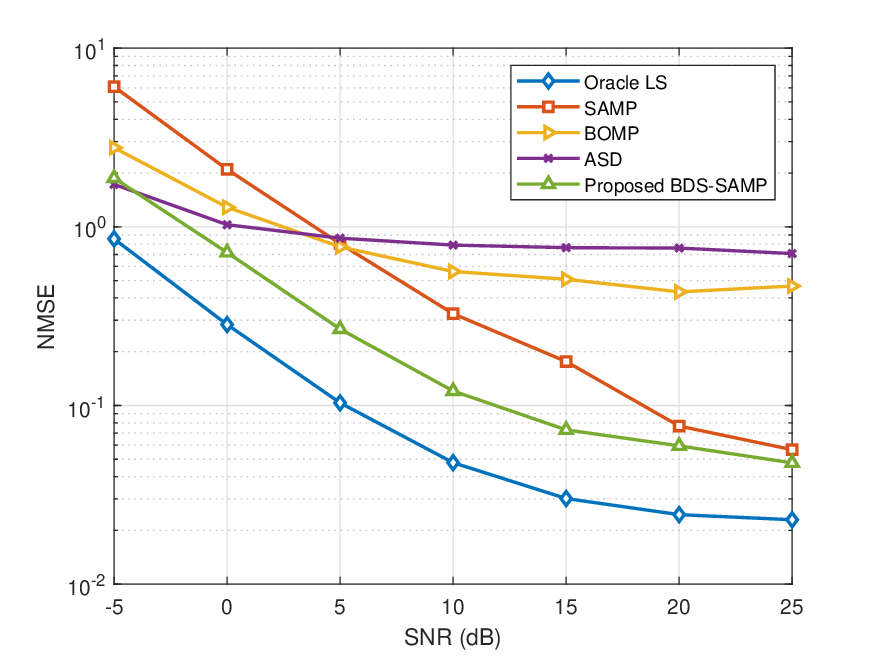}
	\caption{NMSE performance comparison of different channel estimation schemes for SNS massive MIMO systems ($P_h$ = 32, $P_v$ = 32, $K$ = 256, $\rho = 0.45$).}
	\label{fig5}
\end{figure}
\begin{figure}[t]
	\centering\includegraphics[width=3in]{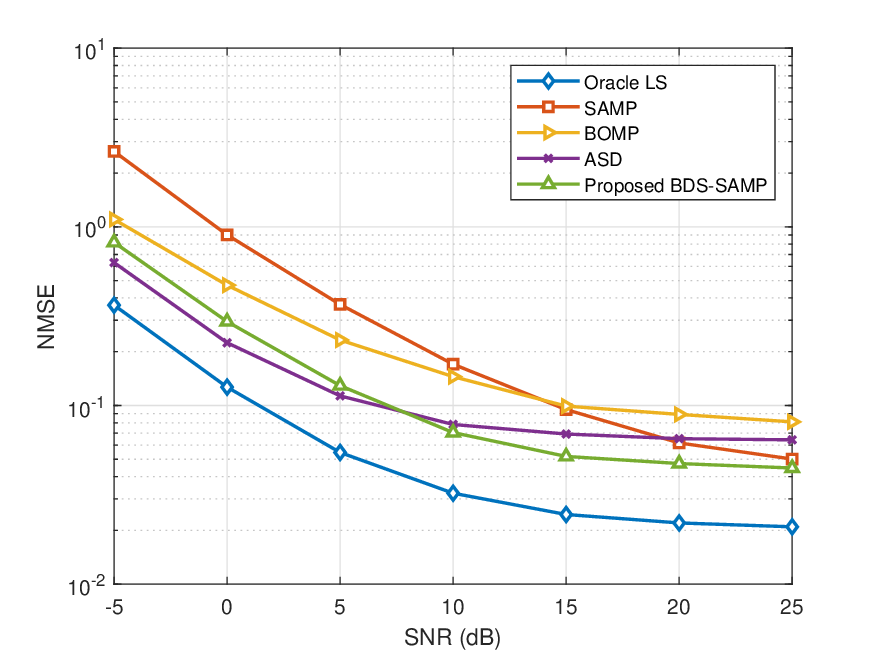}
	\caption{NMSE performance comparison of different channel estimation schemes for SS massive MIMO systems ($P_h$ = 32, $P_v$ = 32, $K$ = 256, $\rho = 0$).}
	\label{fig6}
\end{figure}

Fig. \ref{fig5} shows the NMSE performance of the proposed DBS-SAMP scheme with $P_h$ = 32, $P_v$ = 32, $K$ = 256, and $\rho = 0.45$. For comparison, the results of oracle LS \cite{J19_TSP_XGao_SDEstimation}, SAMP scheme, BOMP scheme \cite{J18_TVT_XMa_BOMP}, and ASD scheme \cite{C16_WCSP_XGao_SDEstimation} are also given.
It can be seen that  the proposed BDS-SAMP scheme can achieve obvious improvement in accuracy compared with SAMP, BOMP, and ASD, especially in high SNR region.
This is mainly due to the adaptively estimation of beam dominant terms on the basis of channel structure and the presetting of power leakage. 
Compared with SAMP, the proposed BDS-SAMP scheme can achieve higher estimation accuracy with the similar complexity. 
At SNR=20 dB, the result of BDS-SAMP scheme has more than 20\%  improvement than that of SAMP.
The oracle LS can directly select the dominant beam supports according to the known sparsity, which is an ideal estimate that can be regarded as the performance upper boundary.
The ASD estimation scheme assumes that the sparsity is known, and the range of the beam domain dominant terms is adjusted in the form of rectangular-blocks. Therefore, the estimation performance of ASD may be better when SNR = -5 dB. 
However, when the number of beams in SNS channel increases, the performance of ASD estimation decreases with the increase of SNR compared with the proposed BDS-SAMP.
Besides, since SNS channels are not regular sparse blocks, simply using BOMP to divide the channel into regular blocks will inevitably lead to poor estimation performance.
Furthermore, since $K$ is much smaller than the dimension of beam domain channel $P$ = 1024, the proposed scheme has low pilot overhead.

Fig. \ref{fig6} shows the beam domain channel estimation performance of the proposed scheme  with $P_h$ = 32, $P_v$~=~32, $K$ = 256, and $\rho = 0$ for SS massive MIMO systems. 
It is clear that the estimation performance of SS channel is generally better than that of SNS channel, but this comes at the cost of channel authenticity. In this case, the proposed BDS-SAMP scheme can perform better in accuracy compared with SAMP, BOMP, and ASD, especially in high SNR region. Besides, the performance of BDS-SAMP is similar to that of ASD in low SNR region. This is mainly because the rectangular-blocks beam structure used by ASD covers the cross-blocks used by BDS-SAMP in the case of low SNR for SS channel.

\begin{figure}[t]
	\centering\includegraphics[width=3in]{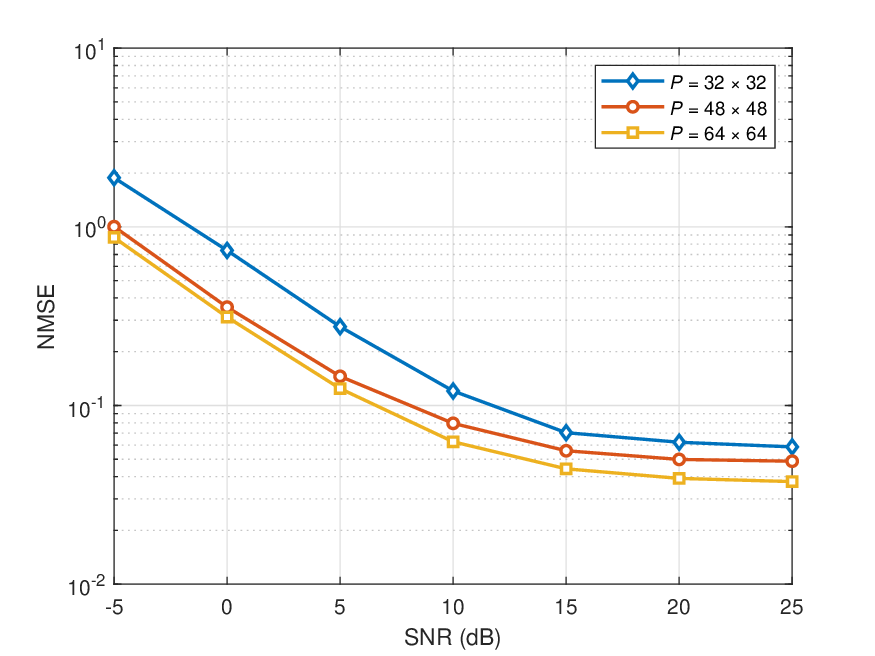}
	\caption{NMSE performance comparison for different sizes of UPA ($K = P/4$, $\rho = 0.45$).}
	\label{fig7}
\end{figure}

Fig. \ref{fig7} displays the NMSE performance of the proposed scheme for different sizes of UPA, i.e., 32$\times$32, 48$\times$48,  and 64$\times$64. It is shown that the performance of NMSE improves with increasing the size of antennas both in low and high SNR regions. 
This is because the power leakage is more severe for larger antenna array with a given VR length.  Fig. \ref{fig5}, Fig. \ref{fig6}, and Fig. \ref{fig7} fully illustrate the accuracy and robustness of the proposed algorithm for massive MIMO channel estimation.

\section{Conclusions}
In this paper, a BDS-SAMP channel estimation scheme for massive MIMO BDCM concerning the spatial non-stationarity and power leakage has been proposed. 
To evaluate our scheme, a realistic massive MIMO BDCM has been adopted to capture spatial non-stationarity.
The sparse structure and power leakage effect have been further demonstrated. 
Based on the cross-block sparse structure and power ratio threshold of beam domain channel, the beam support set has been iteratively obtained in BDS-SAMP channel estimation scheme. 
The simulation results have validated the accuracy and effectiveness of the proposed scheme. 
The proposed BDS-SAMP scheme can achieve obvious improvement in accuracy compared with SAMP, BOMP, and ASD for SNS massive MIMO systems. What’s more, when the number of antennas increases, the proposed scheme can perform well for massive MIMO systems.

\section*{Acknowledgment}
This work was supported by the National Natural Science Foundation of China (NSFC) under Grants 62394290, 62394291, 61960206006, 62301365, and 62271147, the Fundamental Research Funds for the Central Universities under Grant 2242022k60006, the Key Technologies R$\&$D Program of Jiangsu (Prospective and Key Technologies for Industry) under Grants BE2022067, BE2022067-1, and BE2022067-3, the EU H2020 RISE TESTBED2 project under Grant 872172, the High Level Innovation and Entrepreneurial Doctor Introduction Program in Jiangsu under Grant JSSCBS20210082, the Start-up Research Fund of Southeast University under Grant RF1028623029, and the Fundamental Research Funds for the Central Universities under Grant 2242023K5003.



\end{document}